\documentclass{PoS}

\usepackage{latexsym,graphicx,color,amsmath}
\usepackage{amssymb,mathrsfs}
\usepackage{enumitem}
\usepackage{ marvosym }
\usepackage{cite} 
\usepackage{cleveref} 
\usepackage{subcaption} 

\newcommand{\ffr}[2]{\ensuremath{\frac{\displaystyle #1}{\displaystyle #2}}}
\def\nn{\nonumber}
\def\<{\langle} 			
\def\>{\rangle} 
\newcommand\bcc{\begin{pmatrix}}
\newcommand\ecc{\end{pmatrix}}

\title{Prony methods for extracting excited states}

\ShortTitle{Prony methods for extracting excited states}

\author{\speaker{Kimmy K. Cushman}\\
        Yale University\\
        E-mail: \email{kimmy.cushman@yale.edu}}

\author{George T. Fleming \\
        Yale University\\
        E-mail: \email{george.fleming@yale.edu}}

\abstract{We propose an algebraic method for extracting excited states from lattice gauge theory correlation functions. Instead of fitting to a sum of decaying exponentials, we adopt a variant of Prony's method to obtain $M$ energies (exponential decay rates) by finding the roots of an $M^{\rm th}$ order polynomial, and then solving for the amplitudes linearly. The resulting states tend to have overlapping error ellipses, making identification of states ambiguous. This is especially problematic at large Euclidean times where the signal to noise may be low, as well as when many states are extracted. We propose a variation of K-means clustering to identify each extracted state. }

\FullConference{The 36th Annual International Symposium on Lattice Field Theory - LATTICE2018\\
		22-28 July, 2018\\
		Michigan State University, East Lansing, Michigan, USA.}

\begin{document}

\section{Introduction}
Many lattice gauge theory analyses involve excited state extraction. Modeling the first excited state well helps produce a better variational estimate of the ground state, regardless of whether the excited states are the focus of the study. Additionally, the excited state spectrum encodes two particle scattering phase shifts through the Luscher method~\cite{Davoudi:2018wgb,Luscher1, Luscher2}.
By extracting the spectrum of two particles in a box, this formalism provides a connection to the infinite volume continuum of two particle elastic scattering. Also, a robust method for extracting excited states will be valuable to studies of nuclei because their spectrum will be denser compared to hadrons~\cite{nuclear1, nuc_spec}.  Additionally, extracting excited states from confining gauge theories is crucial for beyond the standard model lattice explorations such as composite Higgs or composite dark matter~\cite{Oliver, Appelquist:2018yqe}. 

Excited state extraction often relies on non-linear least squares fitting to decaying exponentials. Finding an acceptable solution can be hard, requiring experienced user input to guide the minimization, particularly when extracting excited states~\cite{Fleming:2009wb, Fleming:2004hs}. Also, plateau finding in effective mass plots with one or more extracted states is inherently subjective. 
We present a black-box method which requires no user input. Instead of fitting to exponentials, it algebraically extracts a fixed number of states proportional to the number of input data points~\cite{Fleming2010}. Contrary to a least squares fit, this black-box method is exactly constrained. Even for a ground state extraction, our method may be preferable because it provides an earlier time approximation to the ground state compared to a typical effective mass plateau. Furthermore, since this algorithm only requires that the signal be decomposed into a sum of exponentials, our method may be useful to any application where such a signal must be analyzed.
\section{State Extraction Method}
In this section we outline the algebraic method for extracting $M$ states from correlation function data. The method can be generalized to a correlation matrix using Matrix Prony (see~\cite{Berkowitz:2019yrf, Berkowitz:2017smo, Beane:2010em}), but here we focus on extraction from a single correlation function in order to address the challenges of overlapping errors. First we set up notation by reviewing the ground state effective mass. We then introduce Prony's method and apply it to bootstrap samples of lattice correlation function data. 
\subsection{Effective Masses}
Consider a two point function (one element of a correlation matrix) of the following form : 
\begin{align}
C_{ij}(t) &= \<0|\mathcal{O}_i^{\dagger}(t) \mathcal{O}_j(0) |0\>\\
&= \sum_{m=0}^{\infty} \<0|\mathcal{O}_i^{\dagger}|E_m\> {\rm e}^{-E_m t}\<E_m|\mathcal{O}_j |0\> \\
\Rightarrow C_{ij}(t) &= \sum_{m=0}^{\infty} a_{ij,m} \,{\rm e}^{-E_m t}\label{corr}.
\end{align}
Note that the sum is invariant under permutation of labels of states $m$. As we will see, this is problematic when the error regions of pairs $(a_m, E_m)$ overlap because the labeling can be ambiguous. However, a simple method to find the ground state energy, $E_0$ is
\begin{align}
 E_0 &= {\rm log}\Big( \ffr{C(t)}{C(t+1)}\Big).
\end{align}
Although this approximation becomes more accurate at larger times, for particles more massive than the pion, the signal to noise ratio decreases exponentially~\cite{Lepage:1989hd, Endres:2011mm}. Therefore, the ground state energy is typically found by first finding an effective mass plateau, but this has inherent systematic bias, and the result will change depending on who selects the plateau.

\subsection{Multiple State Extraction}
To extract M states, one must find the $2M$ parameters $a_m$ and $E_m$ in Eq. \ref{corr}, so one needs a minimum of $2M$ timeslices from the correlation function. The minimal set of consecutive timeslices of the data can be written as a length $2M$ vector $y_n(t)$:
\begin{align}
y_n(t) &\equiv C(t+n) \nn\\
& =  \sum_m^M a_m {\rm e}^{-E_m t} {\rm e}^{-E_m n} \\
\Rightarrow y_n(t) &= \sum_m^M A_m(t) z_m^n,\label{linear}
\end{align}
where we define $A_m(t) = a_m {\rm e}^{-E_m t}$ and $z_m = {\rm e}^{-E_m}$, and the data is modeled as a sum of only the first $M$ states. We can write this as the following matrix equation:
\begin{align}
\bcc y_0 \\ y_1 \\ \vdots \\ y_{2M-1}\ecc & = \bcc 1 & 		1  & \cdots & 1\\
									z_1 &  z_2 & \cdots & z_M\\
									z_1^2 &  z_2^2 & \cdots & z_M^2\\
									\vdots & \vdots & \ddots & \vdots\\
									z_1^{2M-1} &  z_2^{2M-1} & \cdots & z_M^{2M-1}\\ \ecc  \bcc A_1 \\ A_2 \\ \vdots \\ A_M \ecc .\label{vandermonde}
\end{align}
In this algebraic algorithm, we will solve for the $z$'s algebraically with Prony's method and then solve this simple linear problem as usual. Prony's original result~\cite{Prony, phdthesis} can be written in the following way: the non-linear step to find the $z$'s can be done by solving an $M^{\rm th}$ order polynomial from the so-called Hankel matrix determinant
\begin{align}
0 & = \begin{array}{|c c c c c c|}
y_0 		& y_1 	& \cdots &y_{M-2} 	& y_{M-1} 		& 1\\
y_1 		& y_2 	& \cdots & y_{M-1} 	& y_{M} 		& z\\
y_2 		& y_3 	& \cdots &y_M 		& y_{M+1}		& z^2\\
\vdots 	& \vdots 	& \ddots & \vdots 	&\vdots 		& \vdots \\
y_M 		& y_{M+1}& \cdots &y_{2M-2} 	& y_{2M-1}	& z^M
\end{array} \, \,.
\end{align}\\[0.2cm]
 The problem of solving for the energies is then to solve for the roots of a polynomial. 
 This black-box method is exactly constrained and we do not input any initial guesses to the energy and amplitude parameters, as as one must do when fitting to exponentials.

\subsection{Excited State Extraction on Real Data}
\begin{figure}[htp]
  \begin{minipage}[t]{0.30\textwidth}
	 \includegraphics[width = 1\linewidth]{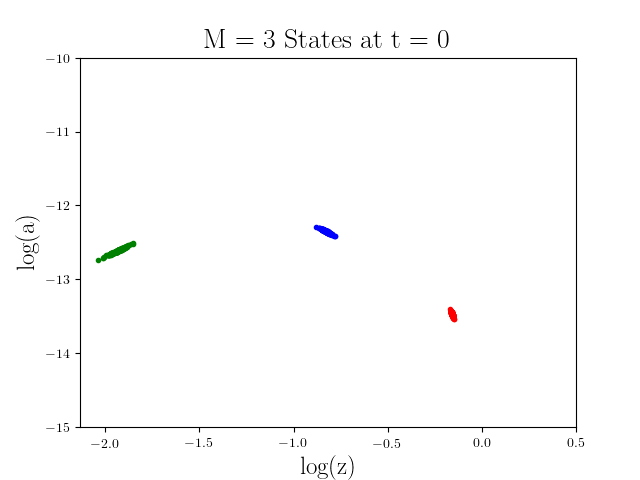}
	 \caption{\small Three state extraction for $y(t=0)$ with well separated error ellipses.}	 
  \label{early_M3}
 \end{minipage}%
\hfill
 \begin{minipage}[t]{0.30\textwidth}
 	\centering
	\includegraphics[width = 1 \linewidth]{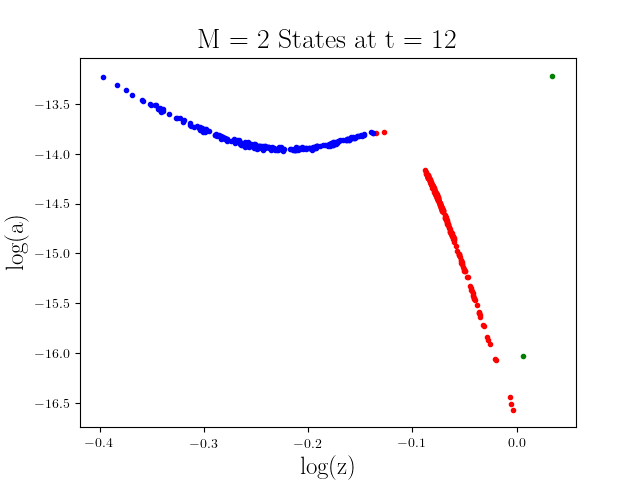}
	\caption{\small Two state extraction for $y(t=12)$ with overlapping error ellipses.}
	\label{M2}
\end{minipage}%
\hfill
 \begin{minipage}[t]{0.30\textwidth}
 	\centering
	\includegraphics[width =1 \linewidth]{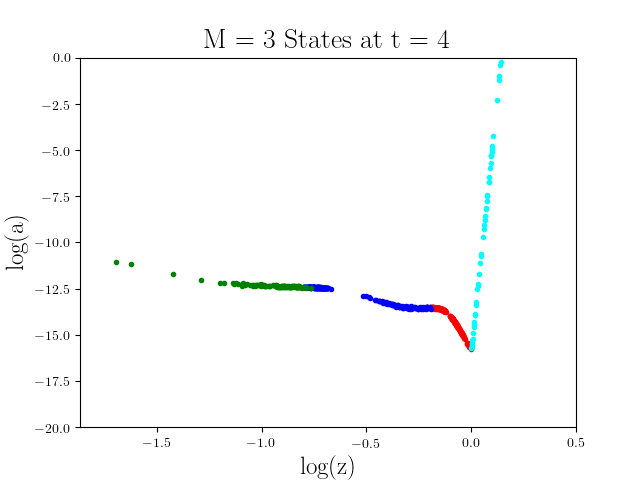}
	\caption{\small Three state extraction for $y(t=4)$ with overlapping error ellipses.}
	\label{M3}
\end{minipage}%
\end{figure}

We show our use of Prony's method on actual correlation function data obtained from the LSD collaboration's investigation of an $SU(3)$ gauge theory with $N_f = 8$ degenerate flavors~\cite{Appelquist:2018yqe}. Figs.~\ref{early_M3},~\ref{M2}, and~\ref{M3} show the extraction of $z_m$ and $a_m$ for two and three state extractions for the pseudoscalar  point-point correlator. Note that the linear Eq. \ref{vandermonde} is solved for $A_m$ after the roots $z_m$ are found. Here we are showing the extraction of the original amplitudes $a_m = A_m {\rm e}^{E_m t}$ from Eq.~\ref{corr}.   

In this extraction, two (three) states were identified from each of 232 bootstrap samples of the $C(t)$ data, and assigned to two (three) distinct states. The state assignment here is determined by the extracted values of $z_m$. The red points are identified as the ground state, and the blue points are identified as the first excited state. The two green points in the two state extraction are labeled as backwards propagating ground states. The green and cyan points in the three state extraction correspond to the second excited and backwards propagating ground states, respectively. These appear to be the correct assignments for $t = 0$ in Fig.~\ref{early_M3} because the distributions do not overlap. However, the signal to noise ratio drops as time goes on,  so the distribution of the extracted parameters $z_m$ and $a_m$ start to overlap (see Figs.~\ref{M2} and~\ref{M3}). Hence, the initial state assignments fail. In order to resolve the best fit parameters $a_m$ and $z_m$ and their errors, we must find the most likely state assignment for each bootstrap sample. 
%
%
\section{Clustering}
In order to re-cluster the points into their most likely states, we will use a variation of the K-means cluster algorithm with expectation maximization. Each of the initial clusters as above is fit to a bivariate Gaussian with probability distribution
\begin{align}
 p(\vec{x}) = \frac{1}{2\pi \sqrt{|\Sigma|}} {\rm e}^{-\frac{1}{2}(\vec{x} - \vec{\mu})^T\Sigma^{-1} 
\,(\vec{x} - \vec{\mu})}
\end{align}
where $\vec{\mu}$ is the mean location of the cluster, and $\Sigma$ is the covariance matrix. Using expectation maximization, a point at $\vec{x}$ would be assigned to cluster 1 instead of cluster 2 if 
\begin{align}
p_1(\vec{x}) &> p_2(\vec{x})\\[0.2cm]
\Rightarrow d_1(\vec{x}) + {\rm log}|\Sigma_1| &< d_2(\vec{x}) + {\rm log}|\Sigma_2|,\label{distance}
\end{align}
where $d(\vec{x}) =  (\vec{x} - \vec{\mu})^T\Sigma^{-1} 
\,(\vec{x} - \vec{\mu})$ is the ``distance'' in standard deviations squared between $\vec{x}$ and $\vec{\mu}$. The log term is a distance penalty for a cluster being larger; a point the same number of standard deviations away from two clusters would be assigned to the smaller cluster. Here, we use a clustering algorithm that looks at each bootstrap sample individually. For $M$ states extracted from the correlations functions, Prony's method provides $M$ points which come from $M$ unique clusters. Therefore our algorithm is as follows:
\begin{enumerate}[topsep = 0pt, noitemsep]
\item Identify initial clusters.
\item Compute the mean and covariance matrix for each cluster.
\item For each bootstrap sample, reassign the $M$ points into $M$ distinct clusters by minimizing the total distance as in Eq.~\ref{distance}.
\item Steps 2 and 3 are repeated until the clusters no longer change.
\end{enumerate}
 \begin{figure}
 \centering
 \begin{minipage}{0.31\textwidth}
 	\centering
	\includegraphics[width = 1 \linewidth]{M3_t4_it0.png}
	\caption{\small Initial cluster assignment as in Fig.~\ref{M3}.}
	\label{M3_0}
\end{minipage}%
\hfill
 \begin{minipage}{0.31\textwidth}
 	\centering
	\includegraphics[width = 1 \linewidth]{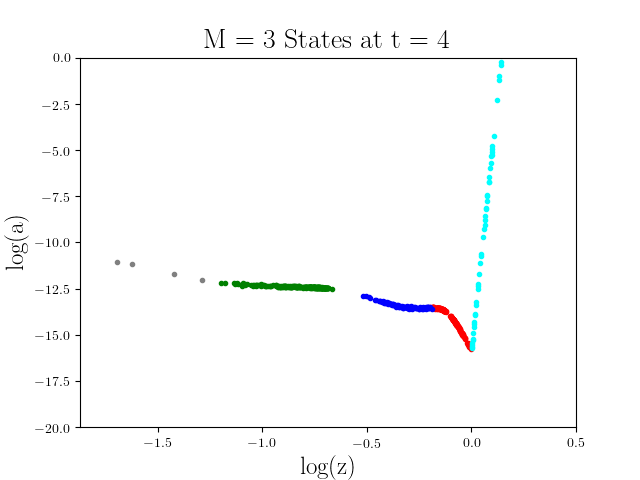}
	\caption{\small After three iterations of clustering}
	\label{M3_3}
\end{minipage}%
\hfill
 \begin{minipage}{0.31\textwidth}
 	\centering
	\includegraphics[width = 1 \linewidth]{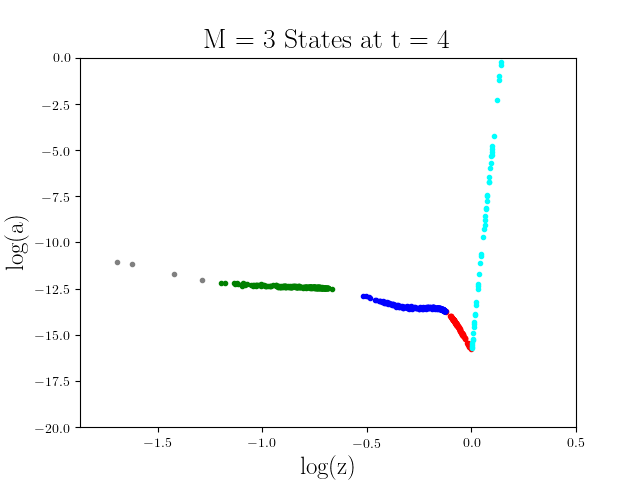}
	\caption{\small After 13 iterations of clustering}
	\label{M3_13}
\end{minipage}%
\end{figure}
Figs.~\ref{M3_0}-\ref{M3_13} show our implementation of the above algorithm on the $M=3, \,t=4$ states. This set of clusters was chosen as an instructive example because the ``correct" cluster assignments can be determined by observation due to the small gaps between clusters. The algorithm appears to recover the clusters correctly. After three iterations, the second excited state in green is separated from the first excited state, and after 13 iterations, the first excited state is separated from the ground state. 

Misestimation of the bivariate Gaussian parameters is highly susceptible to outliers, and the distributions that we are modeling have large tails. In the above cluster re-assignments, we also introduced a ``background cluster" to account for this effect. The probability density to be in the background is uniform and simply the inverse of the area, $A$, of a rectangular region containing the data. In the algorithm, this simply becomes another cluster, and the ``distance" is computed as twice the negative log likelihood as in Eq.~\ref{distance}. That is, $d(\vec{x}) = -2 {\rm log}(1/A)$. The points assigned to the background cluster are interpreted as noise that has a higher probability of being something we are not modeling. The grey points in Figs.~\ref{M3_3} and \ref{M3_13} have been assigned to the background cluster. 
\section{Results and Discussion}
In this section we report the effective masses of the two and three state extractions and compare them to the simpler one-state extraction. We then discuss how the algorithm may be improved. 
\subsection{Preliminary state extractions}
\begin{figure}
 \centering
 \begin{minipage}{0.49\textwidth}
 	\centering
	\includegraphics[width = 0.95 \linewidth]{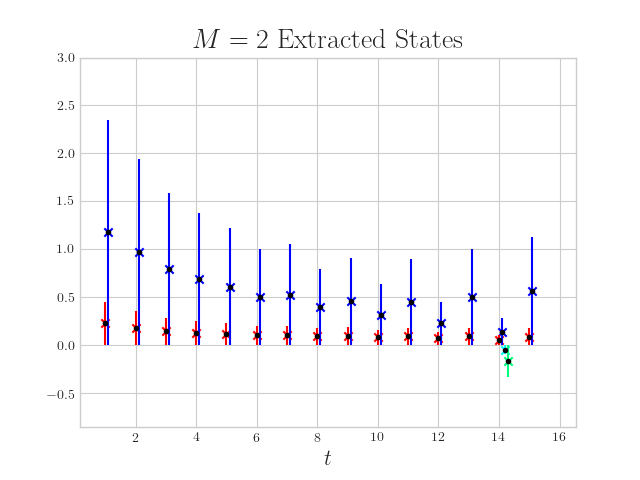}
	\caption{\small Two state extraction results for $y(t=12)$.}
	\label{M2_states}
\end{minipage}%
\hfill
 \begin{minipage}{0.49\textwidth}
 	\centering
	\includegraphics[width = 0.95 \linewidth]{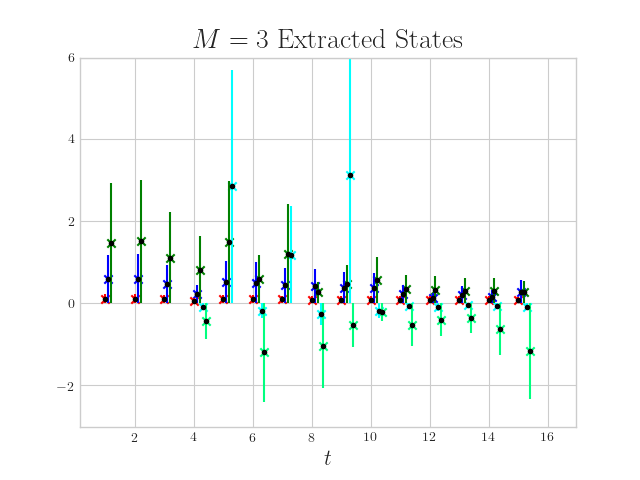}
	\caption{\small Three state extraction results for $y(t=4)$ .}
	\label{M3_states}
\end{minipage}%
\end{figure}

In Figs. \ref{M2_states} and \ref{M3_states} the two and three state effective masses are plotted as a function of time. For $M=2$, the ground state has very little exited state contamination; it approaches its infinite time or true value much earlier than a typical ground state effective mass. The extracted first excited state is absorbing most of the excited state contamination. Similarly, in the $M=3$ extraction, we see most of the excited state contamination has been removed from the first excited state. In both plots we can see the backwards propagating state in cyan.

\begin{figure}
 \centering
 \includegraphics[width = 0.65 \linewidth]{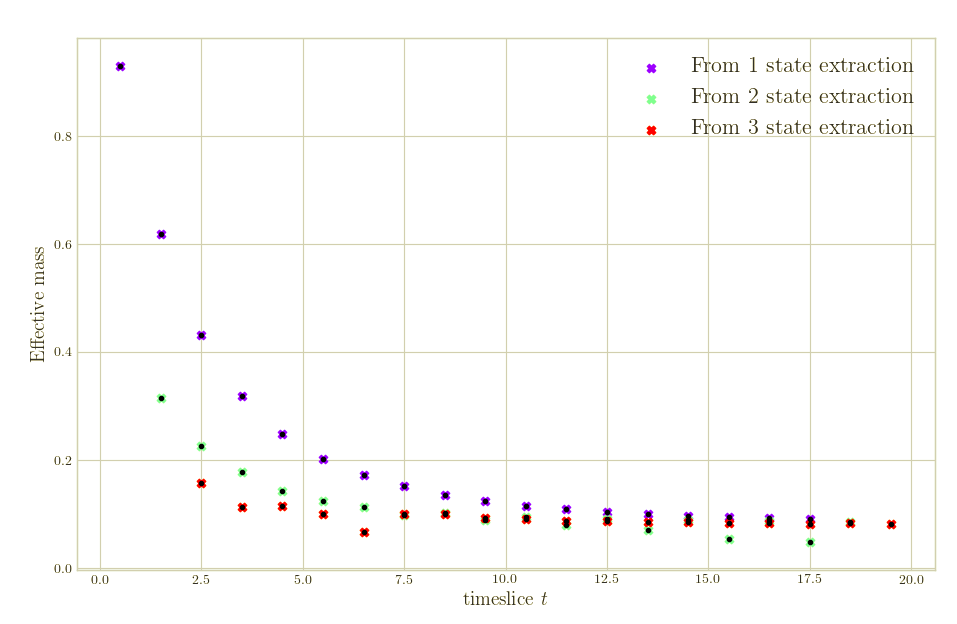}
 \caption{\small Ground state extractions with $M = 1, 2, 3$. Bootstrap errors are smaller than the size of the points.}
  \label{Eff_ground}
 \end{figure}

In Fig. \ref{Eff_ground} we compare the extraction of the ground state as a function of time for different numbers of total extracted states. As expected, the more states we account for, the faster the mass approaches its infinite time or true value. This is because the ground state has less excited state contamination for two and three state extractions, so a reasonable plateau could be assigned to these states at much earlier times as compared to the single state extraction. 

\subsection{Discussion}
One could do studies on fake data. However, using a simple approach to add noise to fake data, we found it was too easy to cluster overlapping distributions of the resulting parameters. Producing fake {\it but realistic} lattice correlation function data is very difficult because lattice noise is highly correlated, so Gaussian clusters do not capture the complexity of real data. 

To improve this method, we would like to account for the fact that these clusters are non-Gaussian. One method we would like to try to account for the non-Gaussian shaped of the clusters is by using the so-called Tukey depth \cite{Tukey} as a measure of distance from a cluster. As a generalization of percentiles to multiple dimensions, the Tukey depth has the benefit of being a non-parametric statistic, so a model for the distribution is not required. The Tukey distance is defined by a curve such that all of its tangent planes have a certain fraction of the data on either side. Tukey states in his original work that this set of planes forms a closed curve in two dimensions. The Tukey depths of Gaussian distributed data would replicate their standard deviation ellipses. 

%
The Tukey distance may be a better metric for this data set. Thus, it may improve the efficiency of the clustering algorithm and increase our ability to extract more states.  The Prony method allows us to extract M states from 2M pieces of data. For $N_t$ timeslices, we should be able to extract up to $N_t/2$ states, and we hope to obtain estimates for several of the first excited states. 

\section{Conclusion}
We have presented an analytic method for extracting excited states from lattice correlation function data which in the absence of noise would yield Prony's exact result. This algorithm has the benefit of being an exact algebraic extraction, rather than being a best-fit search in a high-dimensional parameter space. However, with realistic, noisy correlation function data, to do any kind of error analysis, we must assign a consistent set of label across a set of ensembles, and hence clustering the distributions of parameters is required. With these challenges in mind, this method may still be preferable to conventional approaches because it has the possibility to remove all user input to the process of extracting excited states.  

\section{Acknowledgements}
This material is based upon work that is supported by the Visiting Scholars Award Program of the Universities Research Association. KC acknowledges support from the United States Department of Energy through the Computational Sciences Graduate Fellowship (DOE CSGF) through grant number DE-SC0019323. GF acknowledges support from the United States Department of Energy through grant number DE-SC0019061. We thank the Lawrence Livermore National Laboratory (LLNL) Multiprogrammatic and Institutional Computing program for Grand Challenge allocations and time on the LLNL BlueGene/Q supercomputer. We also thank Argonne National Laboratory (ANL) for allocations through the ALCC and INCITE programs. Computations for this work were carried out in part on facilities of the USQCD Collaboration, which are funded by the Office of Science of the U.S. Department of Energy (DOE).

\bibliography{skeleton}
\bibliographystyle{unsrt}

\end{document}